\documentstyle[12pt]{article}
\newcommand{\bea}{\begin{eqnarray}}
\newcommand{\eea}{\end{eqnarray}}
\newcommand{\bear}{\begin{eqnarray*}}
\newcommand{\eear}{\end{eqnarray*}}

\begin{document}
\begin{center}
{\bf NEW INTEGRABLE GENERALIZATION OF THE ONE-DIMENSIONAL t-J MODEL }
\end{center}
\begin{center}
{F.C.Alcaraz and R.Z.Bariev\footnote{Permanent address: The Kazan Physico-Technical Institute of the Russian Academy of Sciences, 
Kazan 420029, Russia}}
\end {center}
\begin{center}
{Departamento de F\'{\i}sica, 
Universidade Federal de S\~ao Carlos, 13565-905, S\~ao Carlos, SP
Brasil}

\end{center}
\vspace{1cm}
\begin{center}

\vspace{1cm}

\today
\end{center}
\begin{abstract}
A new generalization of the t-J model with a 
nearest-neigbour hopping is formulated and solved 
exactly by the Bethe-ansatz method in the thermodymanic limit.
The model describes the dymanics of fermions with different spins and with
isotropic and anisotropic interactions. 

\end{abstract}
\newpage
In recent years there has been considerable interest in studying low-
dimensional electronic models of strong correlation due to the possilility 
that the normal state of the two-dimensional novel  superconductivity may 
share some interesting features of a 1D interacting electron system [1]. 
In one dimension, the Bethe-ansatz technique can allow one to exactly solve 
Hamiltonians in special cases, such as the Hubbard model [2] and the
ordinary $t-J$ model at its supersymmetric point [3,4]. For example, it is
possible to obtain the low-energy gapless excitation spectrum around the
ground state by the finite-size scaling method [5,6] and calculate the critical 
exponents of the correlation functions [7-9].

The $t-J$ model is a lattice model on the restricted electronic Hilbert space,
where  the occurrence of two electrons on the same lattice site is forbidden.
This restriction corresponds to an implicitly infinite on-site Coulomb 
repulsion. Two types of interactions between electrons on nearest-neighbours sites are considered: a charge interaction of strength $V$ and a spin-
exchange interaction $J$. The Hamiltonian of the extended version of the
$t-J$ model has the form [3,4,10]
\bea
H &=& - \sum_{j=1}^L \sum_{\alpha =1}^{N} P\left(c_{j,\alpha}^+c_{j+1,\alpha} + 
c_{j+1,\alpha}^+ c_{j,\alpha} \right)P \nonumber\\
&-& \sum_{j=1}^L \left[J\sum_{\alpha\ne\beta}^{N} 
c_{j,\alpha}^+ c_{j,\beta} 
c_{j+1,\beta}^+ c_{j+1,\alpha} + \sum_{\alpha,\beta =1}^{N}
V_{\alpha \beta}c_{j,\alpha}^+ c_{j,\alpha} 
c_{j+1,\beta}^+ c_{j+1,\beta} 
\right]
\eea
where $c_{j,\alpha}$ annihilates an electron with a spin component $\alpha$,
P is the projector on the subspace of non-doubly occupied states, and
$L$ is the lattice size. We have
introduced an anisotropy in the charge interactions through a matrix
$V_{\alpha \beta}$.

In the isotropic case $V_{\alpha \beta} = V$ the Hamiltonian (1) corresponds
to the traditional $t-J$ model which was exactly solved by the Bethe-ansatz method at the supersymmetric point $(V = -J = 1)$ for the case  $S = 1/2$.
[3,4,9-11].
The generalization of this result for the arbitrary spin $S$ was carried
out in [12-14]. Other generalizations of the $t-J$ model were studied in 
[15 -18].
In particular, in [18] the anisotropic generalization of the $t-J$ model
has been constructed and it was shown that the model (1) is solvable for
the arbitrary spin and special values of the coupling $J$ and 
$V_{\alpha \beta}$: 
\bea
J &=& \epsilon_0, \nonumber\\
V_{\alpha\beta} &=& \epsilon_0 [(1+\varepsilon_{\alpha})\cosh\gamma \cdot
\delta_{\alpha\beta} + \exp[sign(\alpha -\beta)\gamma]
(1-\delta_{\alpha\beta})]
\eea
where $\gamma > 0$ is a measure of the anisotropy. It was shown more exactly in [18] that the Hamiltonian (1,2) is the quantum counterpart of the so called Perk-
Schultz model [19] which was diagonalized by Schultz in the most general form
[20] (see also [21-22]). In fact, in paper [19] Perk and Schultz have considered two models, the Hamiltonian of the first one is given by (1,2) and the Hamiltonian of
the second one is given by [19,22]
\bea
H = -\sum_{j=1}^L\sum_{\alpha,\beta =1}^N c_{j\alpha}^+c_{j\beta}
c_{j+1,\alpha}^+c_{j+1,\beta},
\eea
The systems which are described by (3) and their different modifications were
studied in spin formulation in [23-25] (see also references therein). In this letter we present a new set of 
models of strongly-correlated particles which are exactly solvable. In the 
system there are different types of particles and interactions between two 
particles in the neighbour sites are given by either Hamiltonian (1) or
Hamiltonian (3) and depend on the type of particles. Thus we will consider
the system the Hamiltonian of which contains the interactions of both types 
(1) and (3) .

Firstly let us formulate the problem and write down the Hamiltonian. Consider 
a periodic one-dimensional lattice of $L$ sites. Place $n$ particles on this
lattice and specify its nature. We assume $P$ components, denoted
$Q_1,Q_2,...,Q_P$ and define $n_{Q_i}$ as the number of particles of component
$Q_i$
\bea
(\sum_{i=1}^P n_{Q_j} = n).\nonumber
\eea
In each set there is $N_{Q_i}$ 
\bea
(\sum_{i=1}^P N_{Q_i} = N).\nonumber
\eea
types of particles, so that the summation 
\bea
\sum_{\alpha\in Q_i}\mbox{means} \sum_{\alpha=N_{Q_1}+...+N_{Q_{i-1}}+1}^
{N_{Q_1}+...+N_{Q_i}}.\nonumber
\eea
We establish the one-to-one correspondence between particles $\alpha
\Leftrightarrow \bar\alpha (\alpha,\bar\alpha\in Q_j)$. So in each 
set there are $q_{Q_j}$ conjugate pairs of particles 
\bea
[(N_{Q_i}+1)/2] \le q_{Q_i} \le N_{Q_i}\nonumber
\eea
where $[N/2]$ means the integer part of the number $N/2$. The dynamics of
above defined particles is described by the Hamiltonian
\bea
H &=& - \sum_{j=1}^L \sum_{\alpha =1}^{N} P\left(c_{j,\alpha}^+c_{j+1,\alpha} + 
c_{j+1,\alpha}^+ c_{j,\alpha} \right)P \nonumber\\
&-&\sum_{j,i}\sum_{\alpha,\beta\in Q_i}\epsilon_0\varepsilon_i
\lbrack U_{\alpha}^{(i)}U_{\beta}^{(i)}c_{j\alpha}^+c_{j\beta}
c_{j+1,\bar\alpha}^+c_{j+1,\bar\beta}- (1 + \varepsilon_i)
\cosh\gamma n_{j\alpha}n_{j+1,\beta}\rbrack\nonumber\\
&-&\sum_{j,i\ne k}\sum_{\alpha\in Q_i, \beta\in Q_k}\epsilon_0
\lbrace g_{ik}c_{j\alpha}^+c_{j\beta}c_{j+1,\beta}^+c_{j+1,\alpha}
-\exp[sign(\alpha -\beta)\gamma]n_{j\alpha}n_{j+1,\beta}\rbrace,\nonumber\\
\eea
where $\varepsilon_i, \epsilon_0 =\pm 1$and the parameters 
$U_{\alpha}^{(i)}$ play the role of anisotropies inside of each set $Q_i$
and  satisfy
\bea
 U_{\alpha}^{(i)} = 1/U_{\bar \alpha}^{(i)}, \hspace{0.5cm} \sum_{\alpha\in Q_i}
(U_{\alpha}^{(i)})^2 = 2\cosh\gamma
\eea
$g_{ij} = g_{ji}^{-1}$.

The ordinary $t-J$ model given in (1,2) is obtained by choosing sector of
(4) in which there are particles only of one type from each set $Q_j$ and
for this type $\alpha \ne \bar\alpha$. The anisotropic version of the Perk-Schultz model of the second type (3) is obtained by choosing a sector in which
we have particles only of one component, for example, $Q_1$ and $N= L$.
Previously we solved exactly the model (4) at this case also for $N < L$
[25,26] for the different choice of conjugate pairs.
The model which is described by the Hamiltonian (4) is also the generalization of the model considered by Sutherland [10] for the case when particles of the
a given component are not identical and the anisotropy has been introduced in the system. In the general case the Hamiltonian (4) describes the dinamics of fermions with different spins. For example, the case $N_{Q_{i}}= 2$ corresponds 
to $S =1/2$ , the case $N_{Q_{i}}= 3 $ corresponds to $S = 1$ with biquadratic
interactions [26]. For the general case $S_i = (N_{Q_j} -1)/2$ the magnetic interactions inside of each set can be written as a polynomial of degree
$2S_i$ in the spin operator. Moreover the interactions between particles
of different sets also exist.

The exact solution for the eigenstates and eigenvalues of the Hamiltonian (4)
can be obtained within the framework of the Bethe-ansatz method [27,28]
The central object of this method is the two-particle scattering matrix $S$
which is calculated from the single- and two-particle processes described by the
Hamiltonian (4). The nonvanishing elements of the S-matrix are
\bea
S_{\alpha'\beta'}^{\alpha\beta}(k_1,k_2) &=& [\sin(i\gamma-\lambda_1 
+\lambda_2)]^{-1}\hat S_{\alpha'\beta'}^{\alpha\beta}(\lambda_1-\lambda_2);\nonumber\\
\hat S_{\alpha'\beta'}^{\alpha\beta}(\lambda) &=&\delta_{\alpha\beta'}
\delta_{\beta,\alpha'}\sin(i\gamma+\varepsilon_i\lambda)\nonumber\\
&-&\varepsilon_i\delta_{\alpha\bar\beta}\delta_{\beta'\bar\alpha'}
U_{\alpha}^{(i)}U_{\beta'}^{(i)}\sin\lambda; \mbox{for} \hspace {0.5cm}
\alpha,\beta \in Q_i\nonumber\\
\hat S_{\alpha'\beta'}^{\alpha\beta}(\lambda) &=&i\delta_{\alpha\beta'}
\delta_{\beta,\alpha'}\sinh\gamma\exp[isign(\beta-\alpha)\lambda]\nonumber\\
&-&G_{ik}\delta_{\alpha\alpha'}\delta_{\beta\beta'}\sin\lambda; 
\mbox{for} \hspace {0.5cm}
\alpha\in Q_i; \beta \in Q_k, i\ne k
\eea
where $\lambda_j$ ($j=1,2,...,n$)  are suitable particle rapidities 
related to the momenta $\{k_j\}$ of the electrons by
\bea
k_j=\cases{
\pi-\Theta(\lambda_j;\frac{1}{2}\gamma),& ${\epsilon_0}=-1$,\cr
-\Theta(\lambda_j;\frac{1}{2}\gamma),& ${\epsilon_0}=+1$,\cr}
\eea
with the function $\Theta$  defined by
\bea
\Theta(\lambda;\gamma)=2\arctan\left(\cot\gamma\cdot\tan 
\lambda \right) ;\hspace{1cm}
-\pi<\Theta(\lambda,\gamma)\leq\pi.
\eea
A necessary and sufficient condition for the 
applicability of the Bethe-ansatz method is the Yang-Baxter 
equations [27,29 ]. Up to our knowledge the form of the S-matrix (6) is a new one, therefore it is necessary to check the Yang-Baxter equations. We have checked these equations numerically for the different choice  of parameters $U_{\alpha}^{(i)}$ and the sets
$Q_i$. We may note also that
$\hat S $-matrix (6) can be obtained by the baxterization of the following 
Hamiltonian
\bea
H &=& \sum_{j=1}^{N-1} e_j;\nonumber\\
e_j&=& \cosh\gamma + \sum_{i\ne k}\sum_{\alpha\in Q_i, \beta\in Q_k}
\lbrack G_{ik}E_j^{\alpha\beta}E_{j+1}^{\beta\alpha}-\sinh\gamma sign(\alpha -\beta)E_j^{\alpha\alpha}E_{j+1}^{\beta\beta}\rbrack ,\nonumber\\
&+&\sum_{i}\sum_{\alpha,\beta\in Q_i}\varepsilon_1^{(i)}
\lbrack U_{\alpha}^{(i)}U_{\beta}^{(i)}E_j^{\alpha\beta}
E_{j+1}^{\bar\alpha\bar\beta} - \cosh\gamma E_j^{\alpha\alpha}E_{j+1}^{\beta\beta}\rbrack ;
\eea
where the $N\times N$ matrices $E^{\alpha\beta}$ have elements
\bea
(E^{\alpha\beta})_{\gamma\delta} = \delta_{\alpha\gamma}\delta_{\beta\delta}
\nonumber
\eea
The quantum chain (9) have important property that if $e_j$ satisfies the 
 following Hecke algebra [31]
\bea
e_{j} e_{j\pm 1} e_{j} -  e_{j} &=& e_{j\pm 1} e_{j}  e_{j\pm 1} - 
e_{j\pm 1}\nonumber\\
\lbrack e_{j},e_{k}\rbrack &=& 0, \mbox{for} \vert j-k \vert \ge 2 \nonumber\\
e_{j}^2 &=& 2\cosh\gamma e_{j}
\eea
then the S-matrix (6) satisfies the Yang-Baxter equations and the model (4) is
integrable [31].
We have checked that $e_{j}$ satisfies the equations (10) and thus we have proved the integrability of the model under consideration (4).

The Hamiltonian (4) is diagonalized by a standard procedure by imposing periodic boundary conditions on the Bethe function. These boundary conditions can be expressed in terms of the transfer matrix of the non-uniform models (6)  by using the quantum method of the inverse problem [31,32]. 

The rapidities ${\lambda_j}$ that define a n-particle wave function are obtained by solving the equations
\bea
\left[\frac{\sinh(\lambda_j-i\gamma/2)}{\sinh(\lambda_j+
i\gamma/2)}\right]^L=(-1)^{n-1}\Lambda(\lambda_j),
\eea
where $\Lambda(\lambda)$ is the eigenvalue of the transfer matrix
\bea
T_{\{\alpha_l'\}}^{\{\alpha_l\}}(\lambda)=\sum_{\{\beta_l\}}\prod_{l=1}^n
S_{\alpha_l'\beta_l}^{\alpha_l\beta_{l+1}}(
\lambda_l-\lambda), \;\;\; (\beta_{n+1} = \beta_{1}).
\eea
It is simple to verify that besides the numbers of particles 
in each set $n_{Q_i}$ the numbers of "conjugate" pairs in each set 
are conserved 
quantities in the Hamiltonian (4). Here we denote two conjugate
particles of same set paired if they are consecutive particles and 
have no unpaired particles of this set between them.  

In a general case, the complete diagonalization of the transfer matrix (12)
is not a simple problem even in the simplest special cases.
Therefore here we restrict ourself to consideration of the thermodynamic
limit of the Hamiltonian (4), when the periodic and free boundary conditions
are identical.

Firstly consider model (4) in the sector where we have no pairs of particles.
If we have no pairs of a set $Q_i$ then the first interaction term in the Hamiltonian (4) does not work and all particles of this set are identical
and can be considered as one component of the model. In this way
the general model (4) in this sector can be reduced to the anisotropic 
$t-J$ model with $P$ components [18,20] and the diagonalization of the transfer
matrix of the inhomogeneous model (12) gives the following Bethe-ansatz
equations

\bea
\prod_{j'=1}^{m_{\sigma -1}} \frac{\sin(\lambda_j^{(\sigma)} - 
\lambda_{j'}^{(\sigma -1)} + 
\frac{i}{2}\epsilon_{\sigma}\gamma)}{\sin(\lambda_j^{(\sigma)}-
\lambda_{j'}^
{(\sigma -1)} - \frac{i}{2}\epsilon_{\sigma}\gamma)} =
 -\epsilon_{\sigma}^{n_{\sigma}}\epsilon_{\sigma + 1}^{n_{\sigma +1}}
\prod_{\rho =0}^q(G_{\sigma\rho}G_{\rho\sigma +1})^{n_{\rho}}\nonumber\\
\times \prod_{j'=1}^{m_{\sigma}} \frac{\sin(\lambda_j^{(\sigma)} - 
\lambda_{j'}^{(\sigma )} + 
i\epsilon_{\sigma+1}\gamma)}{\sin(\lambda_j^{(\sigma)}-
\lambda_{j'}^
{(\sigma )} - i\epsilon_{\sigma}\gamma)} \prod_{j'=1}^{m_{\sigma +1}} \frac{\sin(\lambda_j^{(\sigma)} - 
\lambda_{j'}^{(\sigma +1)} - 
\frac{i}{2}\epsilon_{\sigma +1}\gamma)}{\sin(\lambda_j^{(\sigma)}-
\lambda_{j'}^
{(\sigma +1)} + \frac{i}{2}\epsilon_{\sigma +1}\gamma)} ;
\eea
where
\bea
 j&=&1,2,...,m_{\sigma}; \;\;\; \sigma = 0,1,...,q-1.\nonumber\\
\lambda_{j}^{(-1)} &=& 0; \;\;\; \lambda_{j}^{(0)} =\lambda_{j};\nonumber\\
n_j &=& m_{j-1} - m_{j};\;\;\;  m_1 =L;  \;\;\; m_{q} = 0;
;\;\;\ n_o = n
\eea
and
\bea
q = P;\hspace{1cm}\epsilon_0 =1;\;\;\; G_{\sigma\sigma} = G_{0\sigma} = 
G_{\sigma 0} = 1;\nonumber\\
\epsilon_i = 
\varepsilon_i; \;\;\; G_{ik}=g_{ik}; \;\;\; (i,k=1,2,...,P)
\eea
 $n_{j}$ is the number of particles from the set $Q_j$.

The total energy and momentum of the model are given in terms of the particle rapidities $\lambda_j$ in the following form
\bea
E&=&-2\sum_{j=1}^n\cos k_j =
2{\varepsilon\varepsilon_1}\sum_{j=1}^n\left(\cosh\gamma-\frac{\sinh^2\gamma}
{\cosh\gamma-\cos2\lambda_j}\right), \nonumber\\
P&=&\sum_{j=1}^nk(\lambda_j) .
\eea

Consider now the model (4) in the general case when in each sector $Q_j$ we
have $n_j$ particles and $n'{_j}$ "conjugate" pairs of particles. The 
reference state $\Psi_0$ is made up of a state in which there are no "conjugate"
pairs. Examining (4 ) we see that when this Hamiltonian acts on a state,
it looks for "conjugate" pairs and replaces them by a sum over all such
pairs from a given set $Q_j$. The second possible process is the permutation of
two neighbour particles or one particle and one conjugate pair. It is important that both processes do not depend on the number of types of particles in each set 
$N_{Q_J}$ up to the boundary. Thus the Bethe-ansatz equations depend only
on $n_j$ and $n'_j$ .Therefore  we may consider the 
Hamiltonian (4) for the case when all $N_{Q_j}=2$.  Certainly the fact 
that the solution in the general 
case is strongly in line with the solution for the case $N_{Q_j}=2$ is
connected with the Hecke algebra. The two models with open boundary conditions
have the same energy levels with different multiplications if they have 
the same underlying Hecke algebra [30].
Thus we consider the Hamiltonian (4) in the thermodynamic limit when the periodic
and open boundary conditions are equivalent
 for the case when all $N_{Q_j}=2$. The Bethe-anzatz equations
for this case will be valid for the arbitrary choice of $N_{Q_j}$. 
The examinination of the case  $N_{Q_j}=2$ shows that the Hamiltonian ( 4)  
can be reduced
directly to the 2P-component $t-J$ model [18,20]. Thus the general solution
of the model ( 4) is given by (13 -14) and   (16) with redefined parameters 
$\varepsilon_{\sigma}$ and $G_{\sigma\rho}$
\bea
q = 2P; \;\;\; \epsilon_0 =1;\;\;\;
G_{\sigma\sigma} = G_{0\sigma} = G_{\sigma 0} = 1;\nonumber\\
\epsilon_i = \varepsilon_{[(i+1)/2]};\;\;\;
G_{ik} = g_{[(i+1)/2],[(k+1)/2]};\;\;\;[(i+1)/2]\ne [(k+1)/2];\;\;\;\nonumber\\
 (i,k=1,2,...,2P);\nonumber\\
G_{2l-1,2l} = G_{2l,2l-1} = \varepsilon_l; \;\;\; (l=1,2,...,P)
\eea
and now $n_{2l-1}$ and $n_{2l}-n_{2l-1}$ are numbers of "conjugate" pairs and
separate particles of the set $Q_l (l =1,2,...,P)$.

The ground state energy and the excited states of the Hamiltonian (4) can be
calculated in principle by straightforward methods on the base of Bethe-ansatz equations (13-14) and (17), which in the thermodynamic limit can be
written as integral equations. However, the solution of these equtions
 depends strongly on sign functions $\epsilon_{\alpha}$ and the phase 
factors $g_{ik}$ and can be subject of separate considerations for the
different choice of these parameters. 

\newpage
\begin{center}
{\bf Acknowledgments}
\end{center}
This work was supported in part by Conselho Nacional de Desenvolvimento 
Cient\'{\i}fico - CNPq - Brazil, and by Funda\c c\~ao de Amparo \`a Pesquisa 
do Estado de S\~ao Paulo - FAPESP - Brazil.

\newpage
{\bf REFERENCES}
\begin{enumerate}
\item P. W. Anderson, Phys. Rev. Lett. {\bf 65}, 2306 (1990).
\item  E. H. Lieb and F. Y. Wu, Phys. Rev. Lett. {\bf 20}, 1445 (1968).
\item P. Schlottmann, Phys. Rev. B {\bf 36}, 5177 (1987).
\item P. A. Bares and G. Blatter, Phys. Rev. Lett. {\bf 64}, 2567 (1990).
\item J. L. Cardy, Nucl. Phys. B {\bf 270 [FS16]}, 186 (1986).
\item A.G.Izergin, V.E.Korepin and N.Yu.Reshetikhin, J.Phys.A  {\bf 22}, 2616
 (1989).
\item F.Woynarovich, J.Phys A {\bf 22}, 4243 (1989).
\item H.Fram and V.E.Korepin, Phys.Rev.B {\bf 42}, 10553 (1990); 
       {\bf 43}, 15653 (1991).
\item N. Kawakamiand S.K.Yang  Phys.Rev.Lett.{\bf 65}, 2309 (1990).
\item K. Lee, P. Schlottmann, J. Phys. Colloq. {\bf 49} C8 709 (1988).
\item B. Sutherland, Phys. Rev. B {\bf 12} 3795 (1975).
\item S.Sarkar  J.Phys A {\bf 24}, 5775 (1991).
\item P. Schlottmann, J. Phys. C {\bf 4}, 7565 (1992).
\item N. Kawakami, Phys.Rev. B {\bf 47}, 2928 (1993).
\item A.Kl\"umper, A. Schadschneider and J. Zittartz, J.Phys. A  {\bf 24}, L955 (1991).
\item F. H. L. Essler, V. E. Korepin and K. Schoutens, Phys.Rev.Lett. 
 {\bf 68}, 2960 (1992); 70, 73 (1993).
\item A. Forster and  M. Karowski, Nucl. Phys. B {\bf 408 [FS]}, 512 (1993).
\item R. Z. Bariev, J. Phys. A {\bf 27}, 3381 (1994); 
    R. Z. Bariev, A. Kl\"umper, A. Schadschneider and J. Zittartz,
     Z. Phys. B {\bf 96}, 395 (1995).
\item J.H.H.Perk and C.L.Schultz, Phys.Lett.A {\bf 84}, 407 (1981).
\item Schultz C.L., Physica A {\bf 122}, 71 (1983).
\item H.J.de Vega and E.Lopes, Phys.Rev.Lett.  {\bf 67}, 489 (1991).
\item J.H.H.Perk and Schultz C.L., Physica A {\bf 122}, 71 (1983).
\item A.Kl\"umper, Europhys.Lett.{\bf 9}, 815 (1989);\\
\item R. K\"oberle and  A. Lima-Santos, J.Phys. A {\bf 27}, 5409 (1994).
\item  R. Z. Bariev and F. C. Alcaraz, to be published.
\item  R. Z. Bariev, A. Kl\"umper and J. Zittartz, to be published
\item C. N. Yang, Phys. Rev. Lett. {\bf 19}, 1312 (1967).
\item F. H. L. Essler and  V. E .Korepin, Exactly Solvable Models of Strongly
 Correlated Electrons (World Scientific, Singapore, 1994).
\item  R. J. Baxter, Exactly solved models in statistical mechanics
    (Academic Press, New York, 1982).
\item F. C. Alcaraz, R. K\"oberle and A. Lima-Santos, Int. J. Mod. Phys. 
\item L. Takhtadzhyan and  L. D. Faddeev, 
        Russ. Math. Survey {\bf 34}, 11 (1979).
\item  V. E. Korepin, A. G .Izergin and N. M. Bogoliubov, Quantum Inverse 
Scattering Method, Correlation Functions and Algebraic Bethe Ansatz 
(Cambridge University Press, Cambridge, 1993)

\end{enumerate}
\end{document}